\RequirePackage{fix-cm}
\documentclass[smallextended]{svjour3}       
\smartqed  
\usepackage{graphicx}

\begin{document}

\title{A systematic analysis of the XMM-Newton background: IV. Origin of the unfocused and focused components.
}

\titlerunning{Origin of the unfocused and focused XMM-Newton background}        

\author{F. Gastaldello$^{1}$ \and S. Ghizzardi$^{1}$ \and\\ M. Marelli$^{1}$ \and D. Salvetti$^{1}$ \and S. Molendi$^{1}$ \and\\ A. De Luca$^{1}$ \and A. Moretti$^{2}$ \and\\ M. Rossetti$^{1}$ \and A. Tiengo$^{1,3}$
}

\authorrunning{F. Gastaldello et al.} 

\institute{F. Gastaldello 
           \at $^{1}$ INAF-IASF Milano, via E. Bassini 15, I-20133 Milano, Italy\\
           \email{gasta@iasf-milano.inaf.it}         
           \and
$^{2}$ INAF-Osservatorio Astronomico di Brera, via Brera 28, I-20121 Milano, Italy
           \and\\
$^{3}$ Istituto Universitario di Studi Superiori, piazza della Vittoria 15, I-27100 Pavia, Italy            
}

\date{Received: date / Accepted: date}

\maketitle

\begin{abstract}
We show the results obtained in the FP7 European program\\ EXTraS and in the
ESA R\&D ATHENA activity AREMBES aimed at a deeper understanding of the XMM-Newton background to better design the ATHENA mission. Thanks to an analysis 
of the full EPIC archive coupled to the
information obtained by the Radiation Monitor we show the cosmic ray 
origin of the unfocused particle background and its anti-correlation with the
solar activity. We show the first results of the effort to obtain 
informations about the particle component of the soft proton focused background. 
\end{abstract}

\section{Introduction}
\label{intro}
\subsection{The current knowledge of the XMM-Newton background}
\label{subsec:1.1}
The study of sources of diffuse X-ray emission, from e.g. supernova remnants to clusters 
of galaxies, to the cosmic X-ray background, can yield unique insight into a wide diversity 
of astrophysical phenomena, ranging from collisionless shocks and non-equilibrium plasma 
physics to the build-up of super-massive black holes and to the nature of dark matter. 
Such investigations are intrinsically limited by instrumental background noise, which, 
if not properly characterized, may induce large systematics, preventing to draw firm 
conclusions. Indeed, large amounts of data collected by current X-ray observatories 
remain unexploited because of instrumental background issues. This is particular true
for the data collected by the European Photon Imaging Camera (EPIC) instrument 
\cite{Turner.ea:01,Struder.ea:01} on-board the ESA XMM-Newton mission 
\cite{Jansen.ea:01} in 17 years of observations.

The EPIC instrumental background can be separated into particle and electronic noise components. 
The knowledge of these components has been growing thanks to the many efforts involved in 
collecting suitable blank sky fields to be used as template background by the 
XMM-Newton users \cite{Read.ea:03,Carter.ea:07}, the analysis of the XMM-Newton Guest Observer Facility leading
to the XMM-Newton Extended Source Analysis Software \cite{Kuntz.ea:08,Snowden.ea:08}, the efforts
of the XMM-Newton SOC\footnote{http://xmm2.esac.esa.int/docs/documents/GEN-TN-0014.pdf} and the contributions of various research teams, among them our in Milan has been
particularly active on this topic \cite{De-Luca.ea:04,Gastaldello.ea:07*1,Leccardi.ea:08}.
A summary table of the EPIC instrumental background components is available at this link\footnote{www.star.le.ac.uk/~amr30/BG/BGTable.html}.

The detector noise component is important at low energies, mainly below 0.4 keV for what concerns the 
pn and most of the MOS CCDs. For the chips MOS1-4, MOS1-5, MOS2-2 and MOS2-5 anomalous states have been 
identified, characterized by a significant increase of the count rates below 1 keV. 
These anomalous states can be recognized by performing an analysis of the corner data in a 
hardness-ratio-count rate diagnostic diagram \cite{Kuntz.ea:08}.

The properties (temporal behaviour, spectral and spatial distribution) of the signal 
generated by the interaction of particles with the detectors and with the surrounding structure depend on 
the energy of the primary particles themselves.
High energy particles (E $>$ a few MeV) generate a signal which is mostly discarded on board on the basis 
of an upper energy thresholding and of a pattern analysis of the events \cite{Lumb.ea:02}. 
The unrejected part of this signal represents an important component of the EPIC instrumental background. 
Its temporal behavior is driven by the flux of high energy particles, i.e. Galactic Cosmic Rays (GCRs). 
The time scale of its variability is much larger than the length of a typical observation and is related 
to the variability of Cosmic Rays in the Earth environment linked to the 11-yr solar cycle. 
We call this component the unfocused particle background (or Non X-ray Background, NXB).
There are two ways to measure the quiescent NXB in the EPIC detectors: 1) through the analysis
of portions of the detector not exposed to the sky (\emph{outFOV}) and therefore neither sky X-ray photons 
nor soft protons focused by the mirrors are collected there; 2) through the study of the observations with 
the filter wheel in clos<ed position (FWC): in this configuration, a 1mm thick aluminum window prevents 
X-ray photons and soft protons from reaching the detector. The \emph{outFOV} regions offer the advantage 
of a NXB measurement simultaneous with the observation. FWC observations allow to check and eventually 
correct for spatial variations of the NXB spectrum across the detector.
EPIC MOS has been generally preferred for studies of diffuse sources mainly because 
of the relatively small \emph{outFOV} pn detector area and for the higher percentage of Out of Time (OOT) 
events (6.3\% in Full Frame operation mode or 2.3\% in Extended Full Frame operation mode for the pn as 
opposed to 0.35\% for MOS). In fact, owing to the finite CCD transfer time, a minor fraction of in 
FOV events is wrongly assigned to the \emph{outFOV} region as OOT events. Contamination of soft protons in the
unexposed area of the pn detector due to a different camera geometry with respect to MOS is currently under
investigation. However this complication does not prevent the use of this diagnostic for pn 
\cite{Katayama.ea:03,Fraser.ea:14}.

Another instrument on board XMM-Newton registers the total count rate and basic spectral information on the 
background radiation impinging on the spacecraft, the EPIC Radiation Monitor (ERM).
Its main objective is to issue a warning when the intensity of the radiation exceeds a certain level. 
It consists of two detectors, the low energy proton and electron unit (LE) and the high energy particle 
unit (HE). All the units are based on Silicon diodes, which record the energy loss in the material. 
In particular we made use of the counts detected in single event mode (HES0) in the HE which are sensitive 
to protons in the 8-40 MeV range. For a description of the ERM see this link\footnote{http://www.cosmos.esa.int/web/xmm-newton/radmon-details}.

Low energy particles (mainly protons with E $\sim$ a few tens of keV) accelerated in the Earth magnetosphere 
can also reach the detectors as they are focused through the telescope mirrors. Their interactions with the 
CCDs generate events which are indistinguishable from valid X-ray photons and cannot be rejected on-board. 
When a cloud of such particles is encountered by the satellite, a sudden increase of the quiescent count 
rate is observed. These episodes are known as "soft proton flares" because they are believed to involve 
protons of low energy (soft); the time scale is extremely variable, ranging from hundreds of seconds 
to several hours, while the peak count rate can be more than three orders of magnitude higher than the 
quiescent one. The extreme time variability is the fingerprint of this background component, the 
Soft Proton (SP) component (see \cite{Fioretti.ea:16} and references therein). 
A light curve can immediately show the time intervals affected by a 
high background count rate. Such intervals are usually not suitable for scientific analysis unless 
the X-ray source to be studied is extremely bright. They have to be rejected through 
good-time-interval (GTI) filtering, which consists of discarding all of the time intervals having a 
count rate above a selected threshold.

\begin{figure}
\vskip-1.0truecm
  \includegraphics[width=0.75\textwidth]{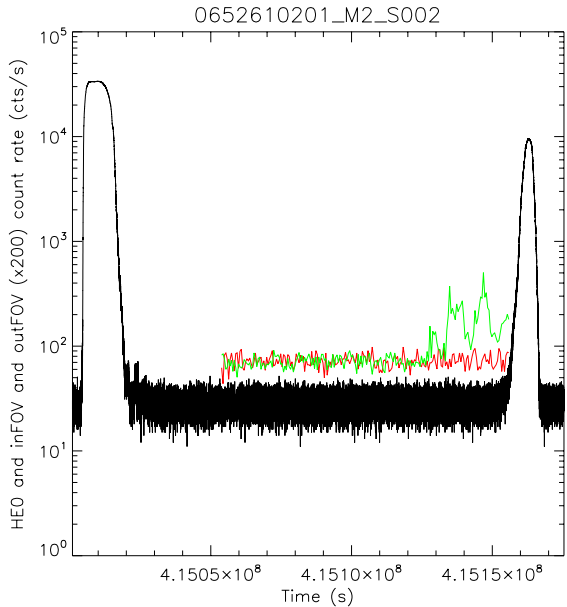}
\caption{ERM HES0 light curve of the rev 2054 (black) together with the EPIC MOS2 light 
curves in the FOV (\emph{inFOV}, green) and outside the FOV (\emph{outFOV}, red) for the 
observation with OBSID 0652610201 (lasting for almost the entire EPIC observation window 
during that orbit), rescaled for display purposes.}
\label{fig:1}       
\end{figure}

\subsection{XMM-Newton as a proton observatory}
\label{subsec:1.2}

An example of a ERM HES0 light curve through a full XMM-Newton orbit is shown in Figure \ref{fig:1} 
together with the light curves of the EPIC MOS 2 count rates within the FOV (\emph{inFOV}) and in the 
unexposed corners (\emph{outFOV}). 
The main features are shown: the high ERM rates at the beginning 
and at the end of the orbit coincide with passage through the Earth radiation
belts, where the EPIC instrument is 
not taking data. The ERM count rate for the rest of the revolution reflects the intensity 
of the Galactic Cosmic Rays (GCRs). The light curve of the outFOV data shows also 
no variation with time, whereas the \emph{inFOV} background rate
is much more variable with flares which are typically not present in the ERM data. The latter 
is the component associated to tens of keV protons concentrated by the mirrors and well 
outside the energy band probed by the ERM.

Therefore XMM-Newton can be considered as a proton observatory covering a wide energy range of these particles:
from the few tens of keV of the soft proton component recorded as the focused flaring background 
component in the EPIC detectors to the tens of MeV protons recorded in the ERM to the hundreds of MeV 
causing the unrejected unfocused component in the EPIC background. It provides a probe of the various
components of the Earth's proton environment: (i) GCRs (ii) solar energetic particles (SEPs) and (iii)
radiation-belt particles \cite{Vainio.ea:09}.
When the two latter components are not present, the normalization of the proton spectrum in the ERM range
reflects only the GCR component. Therefore for example a correlation between the ERM data and the 
\emph{outFOV} data is expected but not yet investigated.

\begin{figure}
  \includegraphics[width=0.65\textwidth]{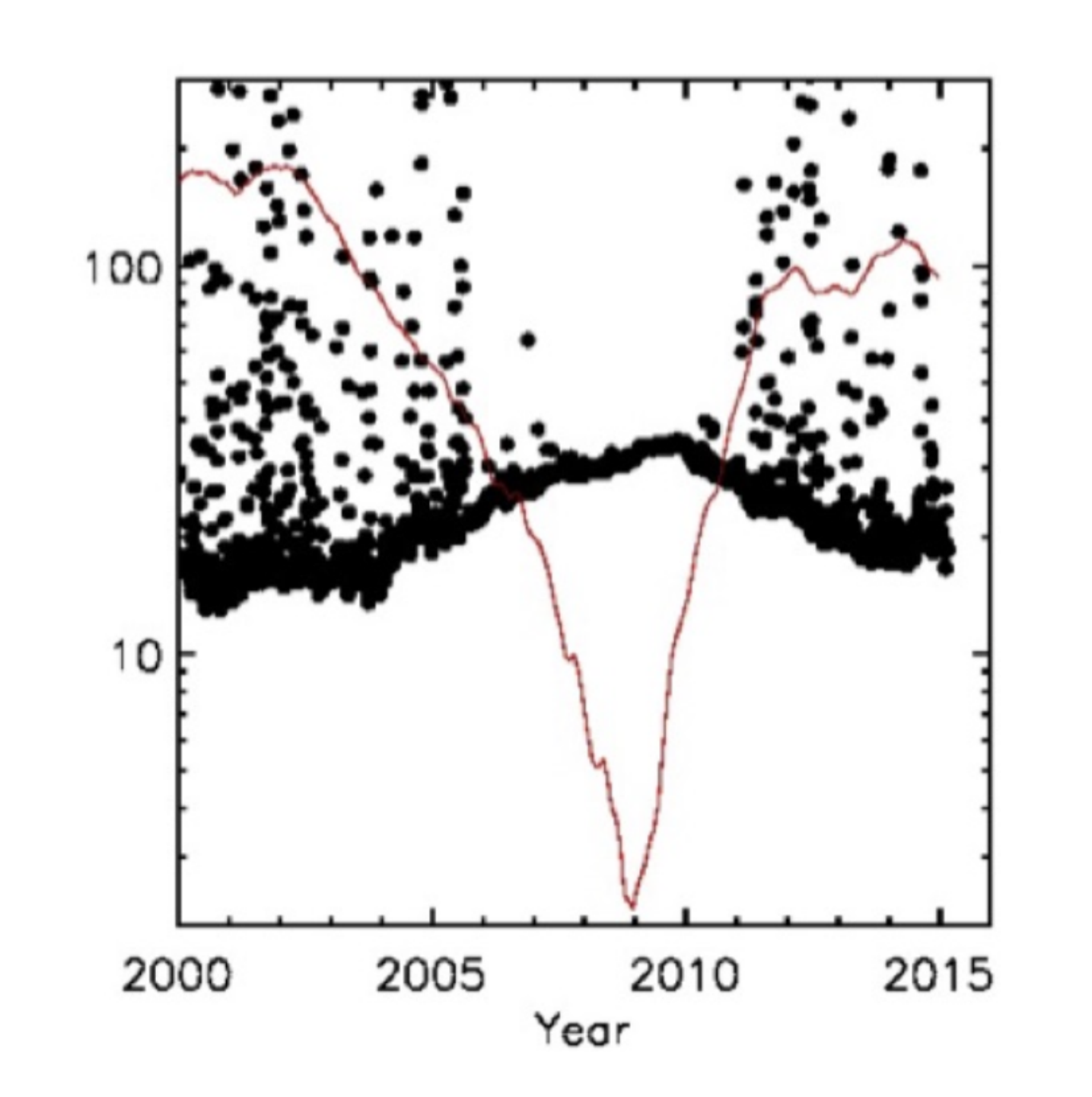}
\caption{The median count rate of the ERM HES0 in each XMM-Newton orbit is shown as a function 
of time with black points. Over-plotted with a red line and in arbitrary units is the number of sun spots taken as a proxy of solar activity. There is a clear general trend of anti-correlation as expected given that most of the time the 8-20 MeV proton flux reflects just the GCR flux. However this is no longer true when SEPs are present which can last for many XMM-Newton orbits. It is also clear from the plot that SEPs are present only during high solar activity.}
\label{fig:2}       
\end{figure}

\subsection{The AREMBES-EXTraS project}
\label{subsec:1.3}

AREMBES (ATHENA Radiation Environment Models and X-Ray Background Effects Simulators) is a R\&D ESA project 
aimed at characterizing the effects of focused and non-focused particles on ATHENA detectors: both in terms 
of contributions to their instrumental background and as source of radiation damage\footnote{http://space-env.esa.int/index.php/news-reader/items/AREMBES.html}. 
XMM-Newton is a test-bed
of the various background components which will be relevant for the ATHENA mission. To this aim
we exploit the entire XMM-Newton public data set to produce the most complete and clean data set ever used 
to characterize the XMM-Newton particle-induced background, taking as input the preliminary results of the FP7
European project EXTraS (Exploring the X-ray Transient and variable Sky\footnote{http://www.extras-fp7.eu/},\cite{De-Luca.ea:15}). 
In order to analyze a dataset as uniform as possible as a function with time we exploited the conservative and stable MOS2 dataset \cite{Marelli.ea:17}.
This paper is part of a series of four describing the data preparation and analysis 
and the scientific results of the project: the first paper set the basic definitions and the methods of the
data analysis \cite{Marelli.ea:17}, the second describes the characterization of the EPIC background 
\cite{Salvetti.ea:17}, the third describes the dependence of the EPIC background with respect to the 
magnetospheric environment encountered by XMM-Newton through its orbit \cite{Ghizzardi.ea:17} and this one
investigates the origin of the focused and unfocused particle background. 

\section{The unfocused particle background}
\label{sec:2}
\subsection{Dependence on solar cycle}
\label{sec:2.1}

\begin{figure}
  \includegraphics[width=0.75\textwidth,angle=-90]{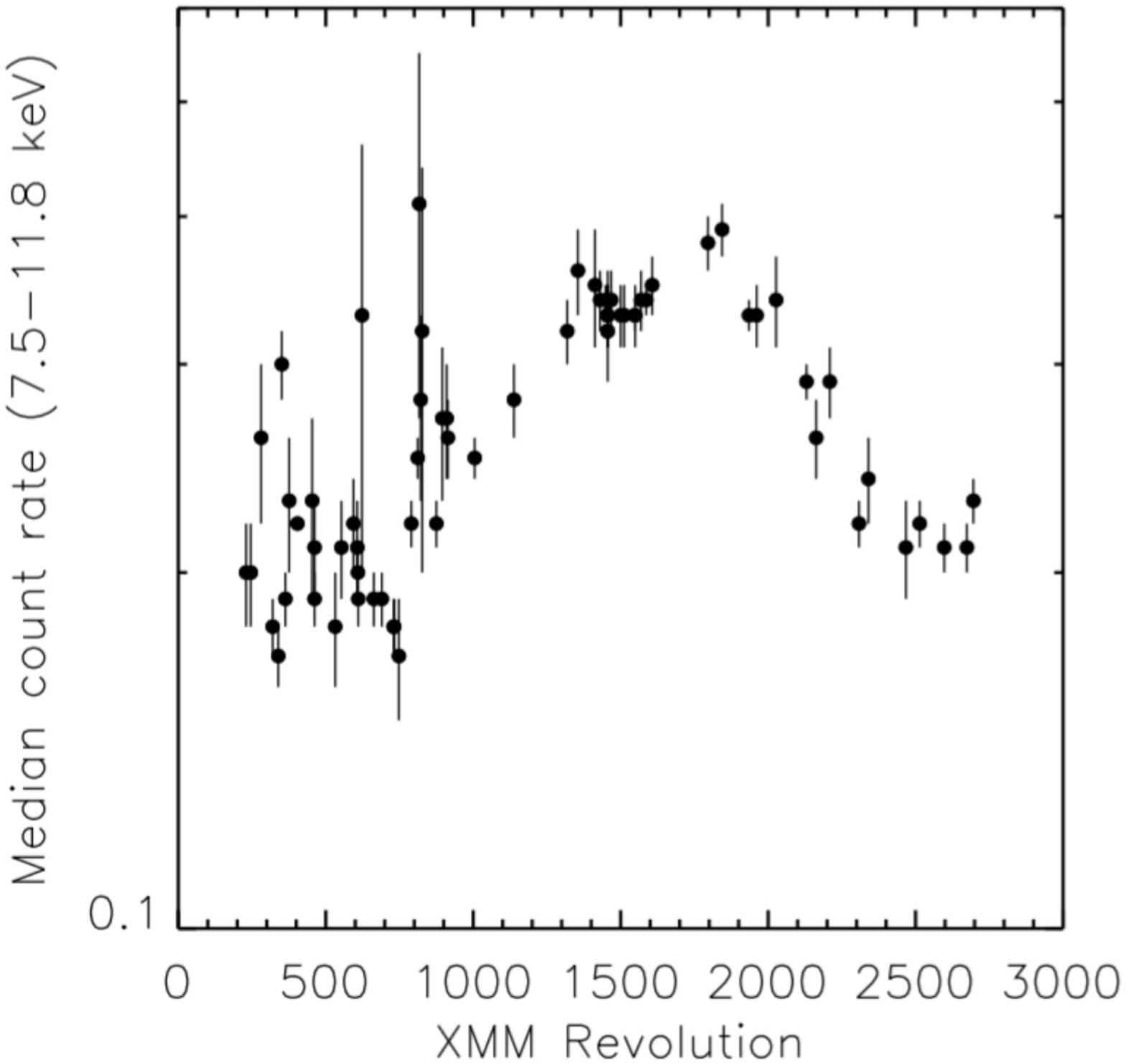}
\caption{Median count rate over the all field of view of the available MOS2 closed observations.}
\label{fig:3}       
\end{figure}

The key temporal variation imposed on the ERM and EPIC data for what concerns the 
unfocused instrumental background is the solar cycle because it modulates the 
Galactic Cosmic Rays. The GCR flux anti-correlates with the solar cycle. 

A useful and easy proxy for the solar activity is the number of sun spots and this is 
plotted aside the median in each XMM-Newton orbit of the ERM HES0 count rate in Figure \ref{fig:2}. 
This plot highlights the fact that the median is effective in removing features due to 
passage in the belts but not periods of enhanced count rates which are due to SEP events. 
These two types of time intervals, passage in the belts 
and SEPs, are periods where the proton flux in the 8-40 MeV range is not just due to GCR.

\begin{figure}
  \includegraphics[width=0.65\textwidth,angle=-90]{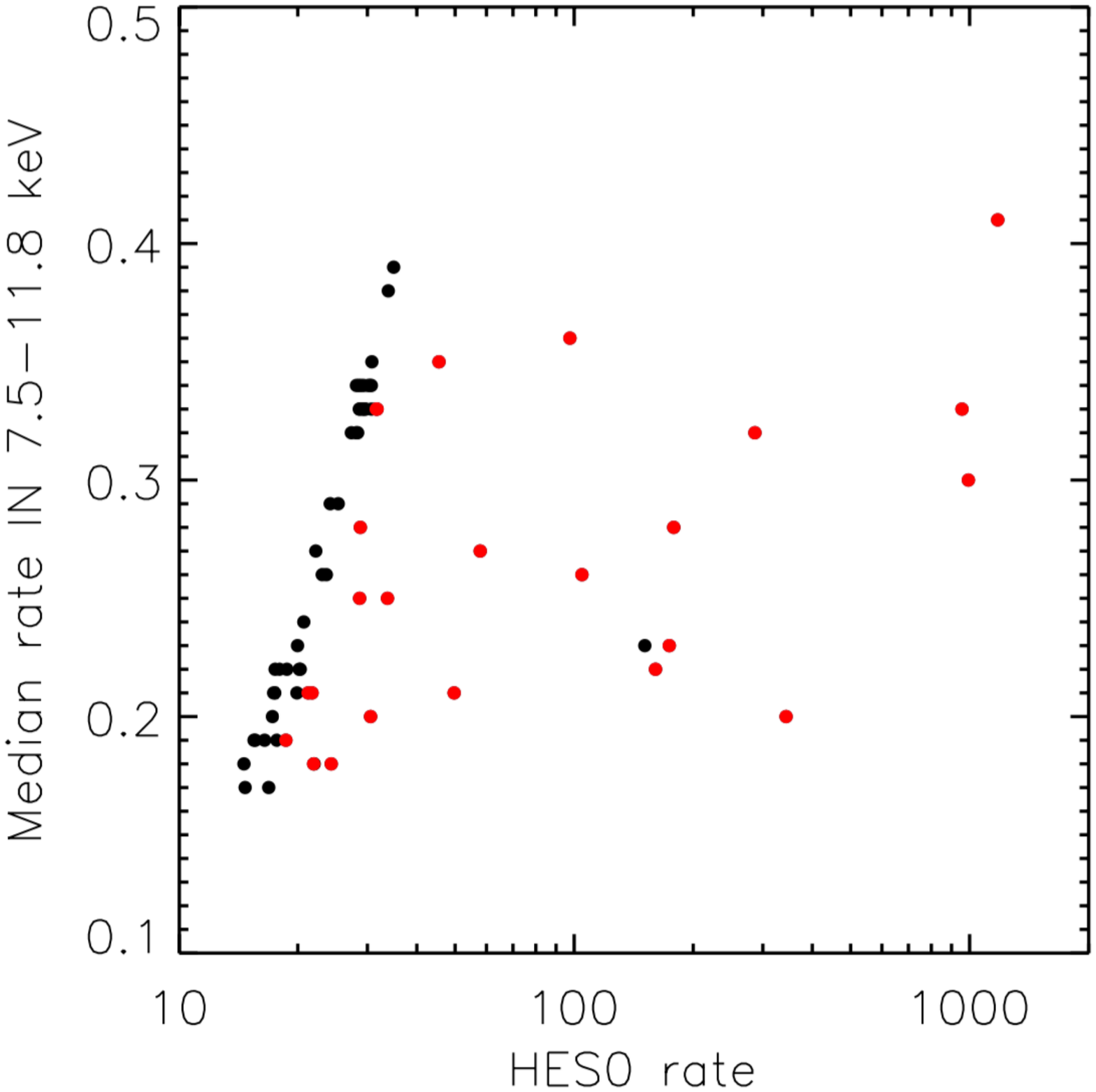}
\caption{Correlation of the ERM HES0 count rate and the corresponding median rate over the 
all \emph{inFOV} for closed observations. The black points show the expected correlation 
when the HES0 count rate is representative of the GCR flux. Red points are selected when 
filtering for SEPs or radiation belts passage as detailed in the following section.}
\label{fig:4}       
\end{figure}

The same temporal behavior is seen when looking at the all set of closed observations 
listed in the XMM-Newton web-site (Figure \ref{fig:3}). Outliers in the relation are due to closed 
observations which are scheduled at the beginning or at the end of the revolution and they 
are therefore affected by high energy particles trapped in the radiation belts.
The key aspect that the instrumental background is correlated with high energy particles 
is also reflected in a naive correlation of the closed data median count rate and the corresponding median ERM HES0 rate during the same time interval (Figure \ref{fig:4}). 
The ERM count rate can vary by up to two orders of magnitude, reflecting the high spectral 
variability of SEPs and particles in the radiation belts, however the instrumental 
background varies at most by a factor of 2.

\subsection{Filtering out SEPs and radiation belts}
\label{sec:2.2}

\begin{figure}
  \includegraphics[width=0.65\textwidth,angle=-90]{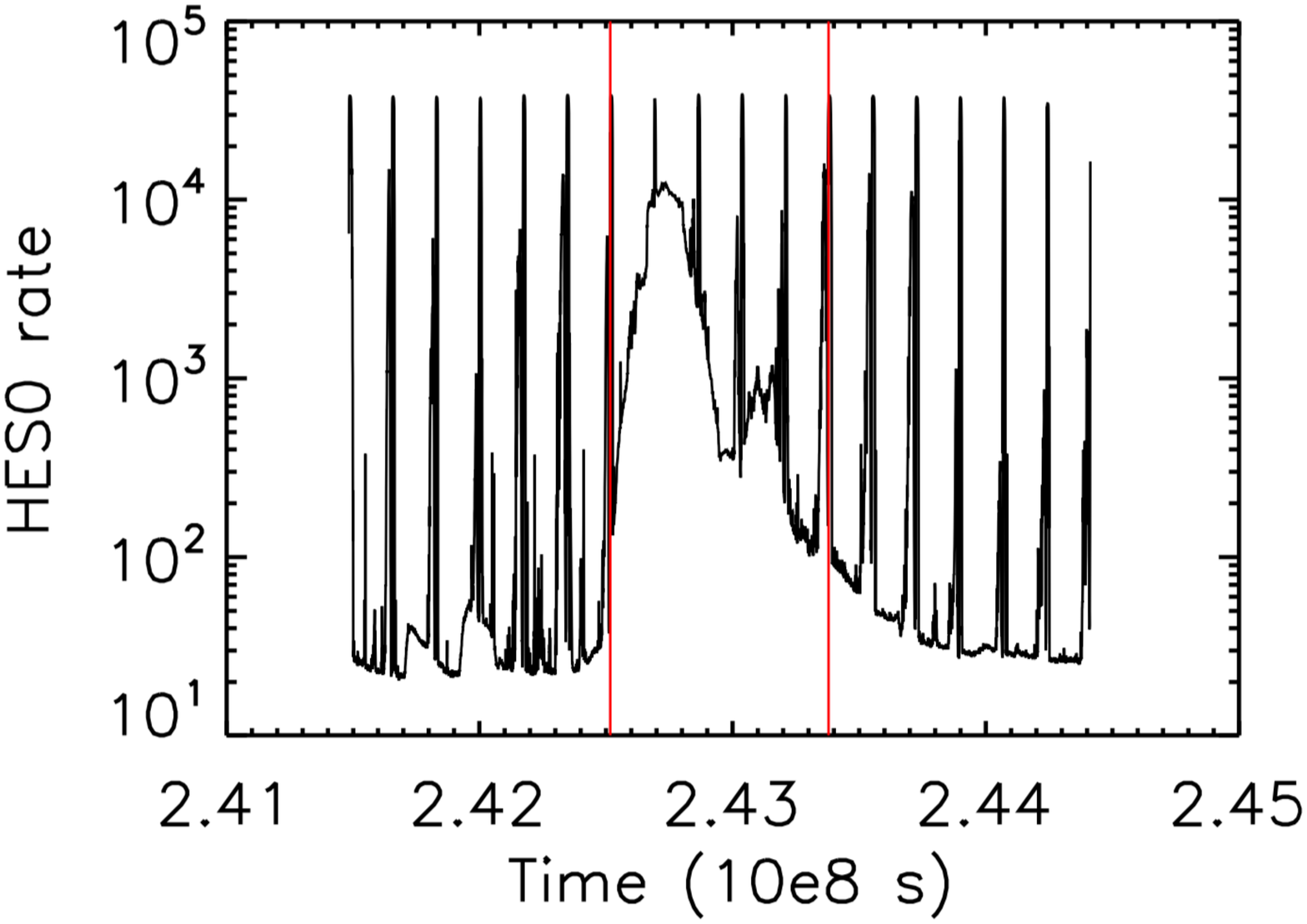}
\caption{ERM HES0 count-rate during several XMM-Newton orbits (from 1056 to 1063) showing the time interval flagged as SEP shown by the vertical red lines. Clearly there is some residual high flux left in the declining tail of the flare.}
\label{fig:5}       
\end{figure}

To obtain a consistent comparison of the count rate in the two instruments is 
therefore necessary to filter periods of radiation belts passage and SEPs events. 
The former is obtained by fitting the histogram of the counts with a Gaussian and 
excluding time periods above $3\sigma$ from the mean, in a similar fashion as filtering soft 
proton flares in the light curves of EPIC observations. The latter has been obtained by 
using the SEP events list found on the ESA Solar Energetic Particle Environment 
Modelling (SEPEM) application server\footnote{http://dev.sepem.oma.be/help/event\_ref.html}. 
The time duration of the SEP event in the list is usually conservative, even though 
sometime this is not true and leads to low residual level of outliers 
(see an example in Figure \ref{fig:5}).

\subsection{Correlation of ERM and \emph{outFOV} MOS2 data}
\label{sec:2.3}

\begin{figure*}
  \includegraphics[width=0.7\textwidth]{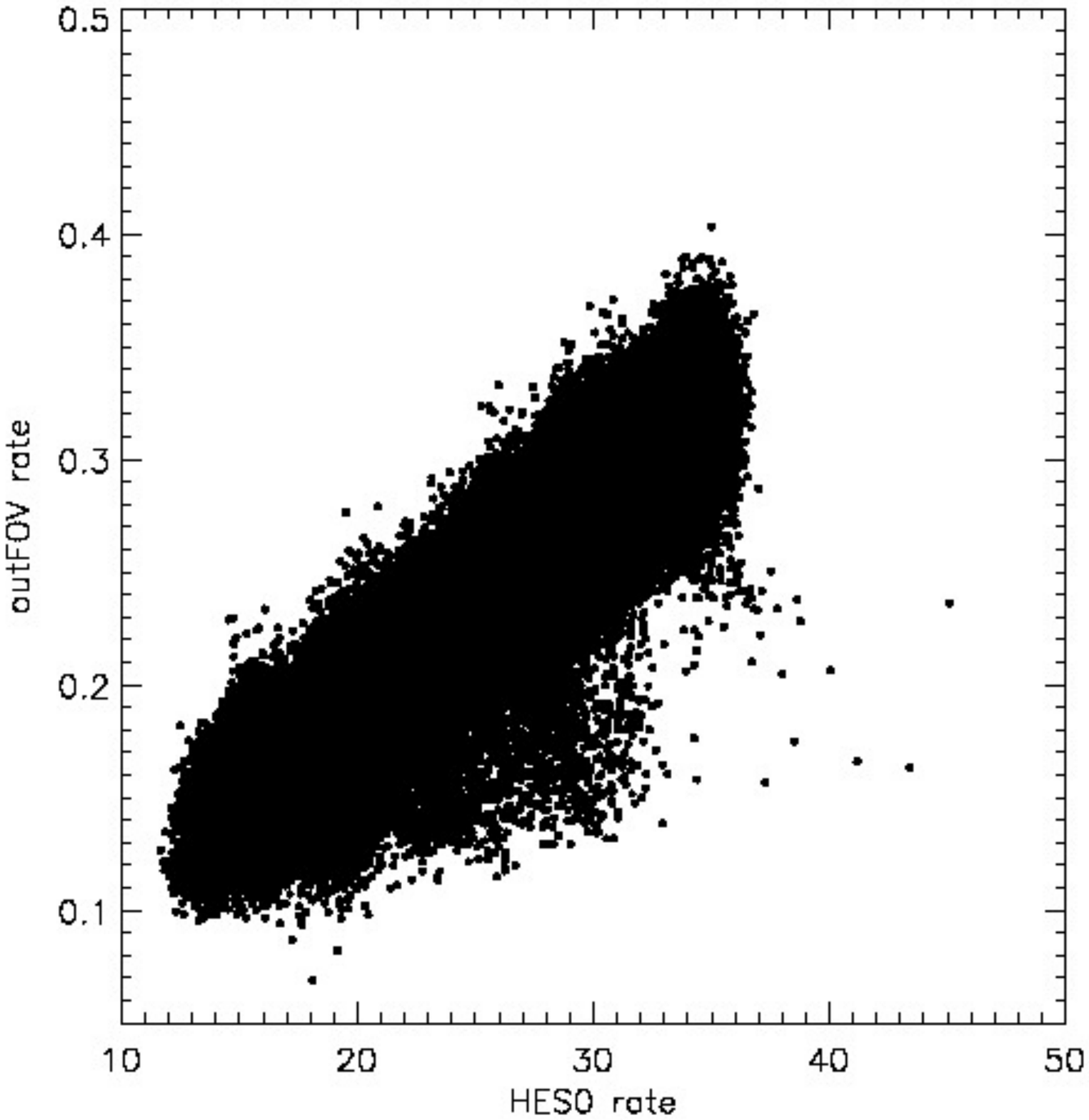}
\vskip -1.0truecm
  \includegraphics[width=0.7\textwidth]{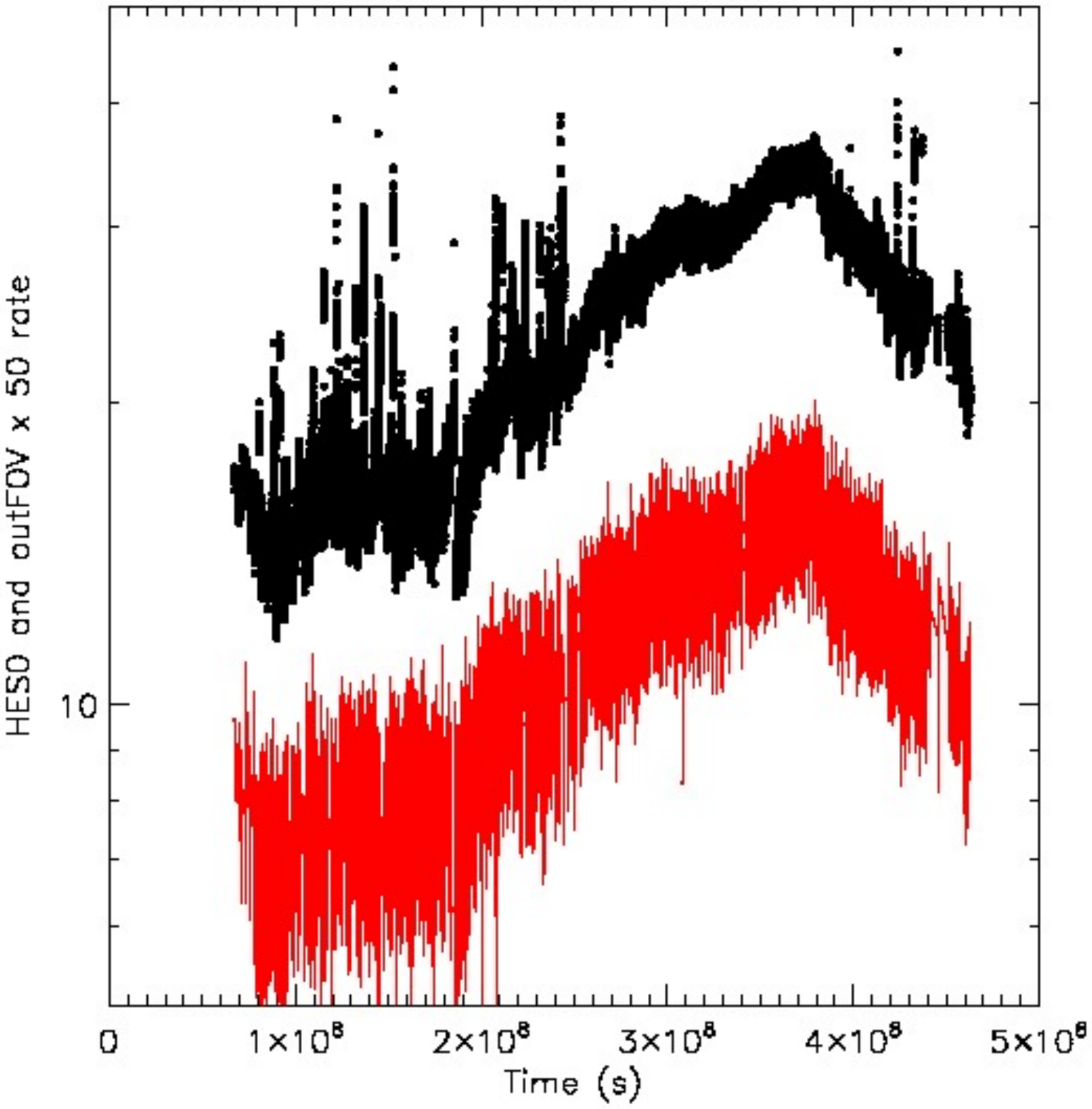} 
\caption{Top panel: plot showing the correlation between ERM HES0 count rates and the corresponding \emph{outFOV} count rate. A clear correlation is present. Bottom panel: time resolved behavior of the ERM HES0 count rate (black) and the EPIC- MOS2 \emph{outFOV} data (red), rescaled for plotting purposes.}
\label{fig:6}       
\end{figure*}

When the ERM data are thus filtered the correlation is evident and also the time behavior
is perfectly consistent, see Figure \ref{fig:6}. The plot corresponds to 71.5 Ms worth of data.
We performed the Spearman and Kendall non-parametric correlation tests which returned values
of the Spearman's $\rho$ of 0.927 and Kendall's $\tau$ of 0.762.

\begin{figure}
  \includegraphics[width=0.55\textwidth,angle=-90]{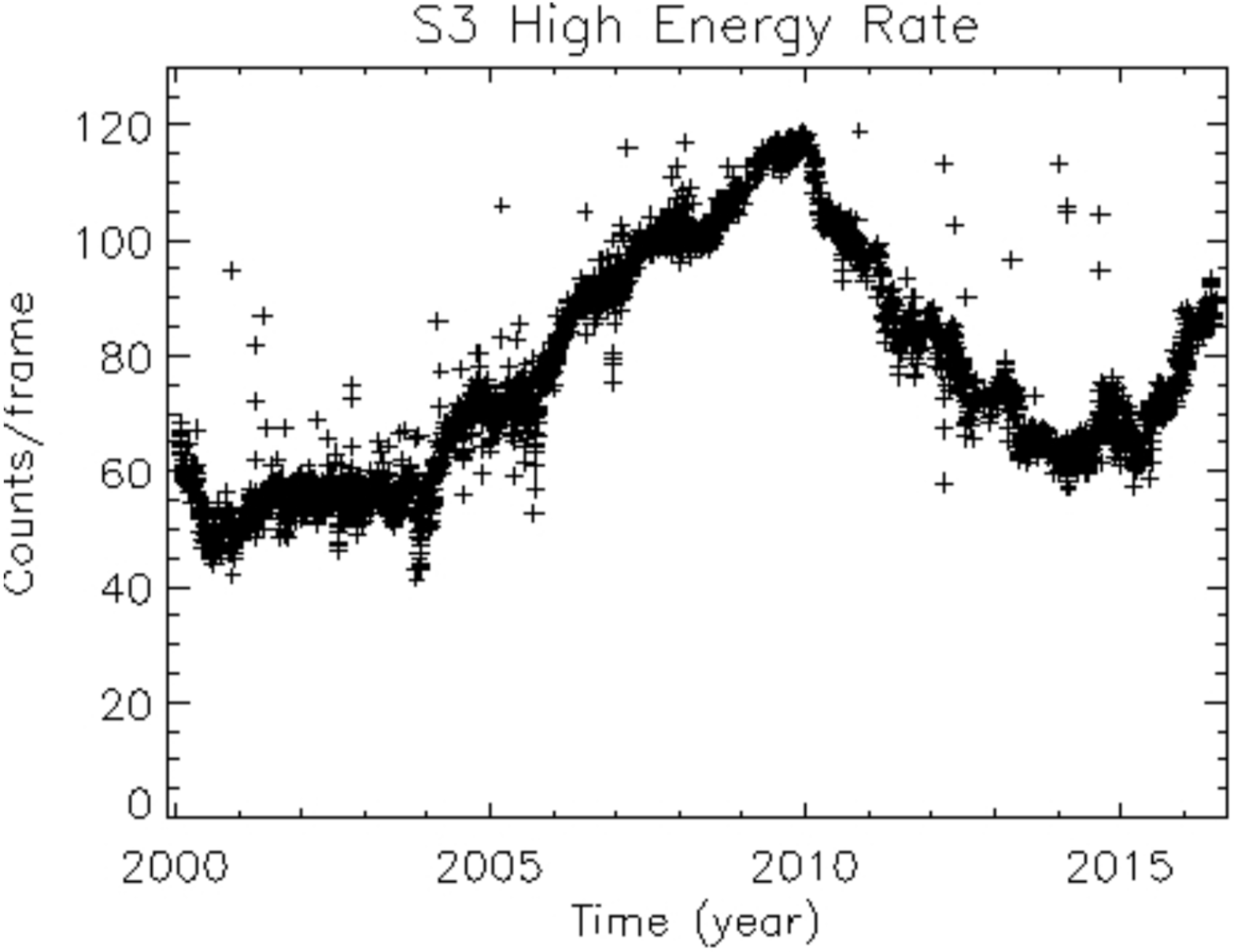}
\caption{Chandra high energy (12-15 keV) count rate for the ACIS-S3 CCD as a function of year.}
\label{fig:7}       
\end{figure}

The same behavior has been found for the Chandra background rate as a function of time, see
Figure \ref{fig:7}, taken from C. Grant web-site$\footnote{http://space.mit.edu/~cgrant/cti/cti120/bkg.pdf}$.
The inference is that the Chandra background is dominated by the GCR rate \cite{Smith.R.K.ea:10}.
The striking similarity reinforces the idea
of a common GCR origin for the unfocused particle background of CCD detectors in similar orbits.

\subsection{Absence of correlation with the magnetospheric environment}

\begin{figure}
  \includegraphics[width=0.65\textwidth]{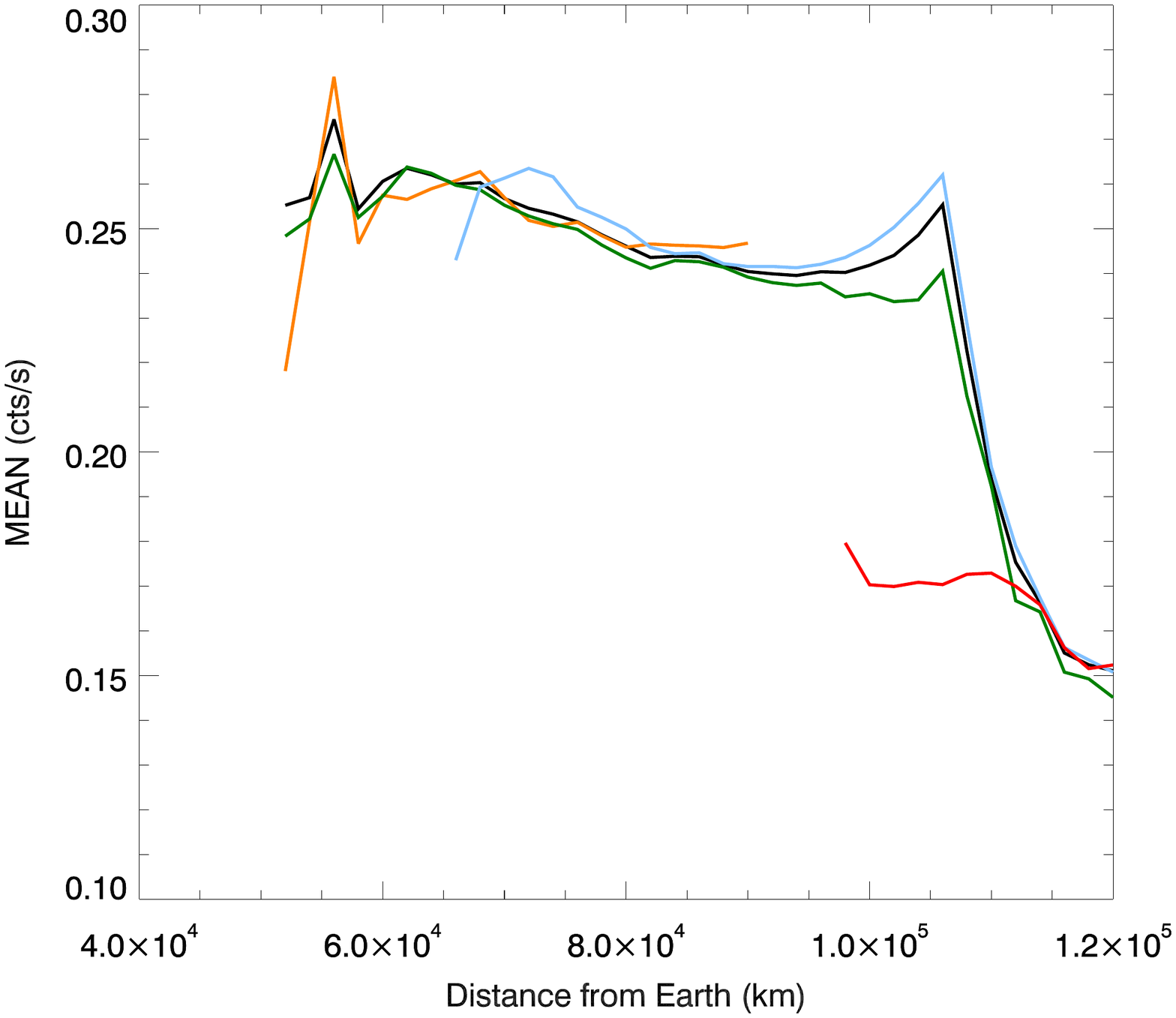}
\vskip -2.0truecm
\caption{Mean count rate as a function of distance from Earth of the outFOV count rate, color coded according to the different magnetospheric regimes defined in \cite{Ghizzardi.ea:17}.}
\label{fig:8}       
\end{figure}

The absence of correlation with magnetospheric environment is yet another evidence of the GCR origin 
of the particle component creating the unfocused particle background in EPIC. The plot shown in Figure 
\ref{fig:8} reports the mean of the \emph{outFOV} rate as a function of the distance from Earth, color coded 
according to the definitions of magnetospheric environments in \cite{Ghizzardi.ea:17}. 
There is no indication of a dependence on the magnetospheric environment: the low rates when the 
XMM-Newton satellite is outside of the bow shock are simply due to the fact that the satellite probed 
this magnetospheric regime at the beginning of the mission, when solar activity was high and therefore 
the GCR flux and its induced particle background was low.

\section{The focused particle background}

\subsection{Data selection}
The objective of this part of the work is the comparison of the XMM-Newton focused background caused by soft protons
with environmental estimates of the soft proton particle flux recorded by orbiting satellites designed
and calibrated to measure those particles, in order to estimate the concentration power of the XMM-Newton optics.
We used as primary datasets the (\emph{inFOV}-\emph{outFOV}) XMM-Newton rate which reflects the intensity of the
soft proton component (when the count rate is above 0.1 cts/s,\cite{Salvetti.ea:17}) 
and the data from the Advanced Composition Explorer
(ACE) satellite in orbit around L1 \cite{Stone.ea:98}, chosen for a time span of available data comparable
to the one we have for XMM-Newton.
We used particle data from the Low Energy Magnetic Spectrometers (LEMS), LEMS120 and LEMS30, of the EPAM 
instrument dedicated to 
monitor the low energy (46 keV - 4.8 MeV) protons \cite{Gold.ea:98}. Of particular interest for our purposes
are the low energy channels of those detectors, P1 which covers the 46-67 keV energy range and P2 which 
covers the 67-115 keV energy range ($^{\prime}$ refers to the channels for LEMS120). 
LEMS30 points at 30$^{\circ}$ from the Sunward pointing spin-axis
 and LEMS120 points at 120$^{\circ}$ from the spin axis, therefore looking back towards the 
Earth's bow-shock. Because of this orientation LEMS120 is sensitive to upstream events (brief, intermittent
particle enhancement) when magnetically connected to the to Earth's bow-shock. The LEMS30 detector with
its different orientation is not as sensitive to upstream events (e.g., \cite{Haggerty.ea:00,Tessein.ea:15}).
Further the LEMS30 P1 channel has no data since day 327 of 2001 and P2 since day 302 of 2003 
\cite{Haggerty.ea:06}. We will therefore base mainly our analysis on the LEMS120 P1$^{\prime}$ and P2$^{\prime}$ 
channels.
We took the 5 minutes average calibrated Level 2 data from the ACE Science Center\footnote{http://www.srl.caltech.edu/ACE/ASC/level2/lvl2DATA\_EPAM.html}. 

\subsection{Comparison of  \emph{inFOV}-\emph{outFOV} MOS2 and ACE EPAM data}

\begin{figure*}
  \includegraphics[width=0.5\textwidth]{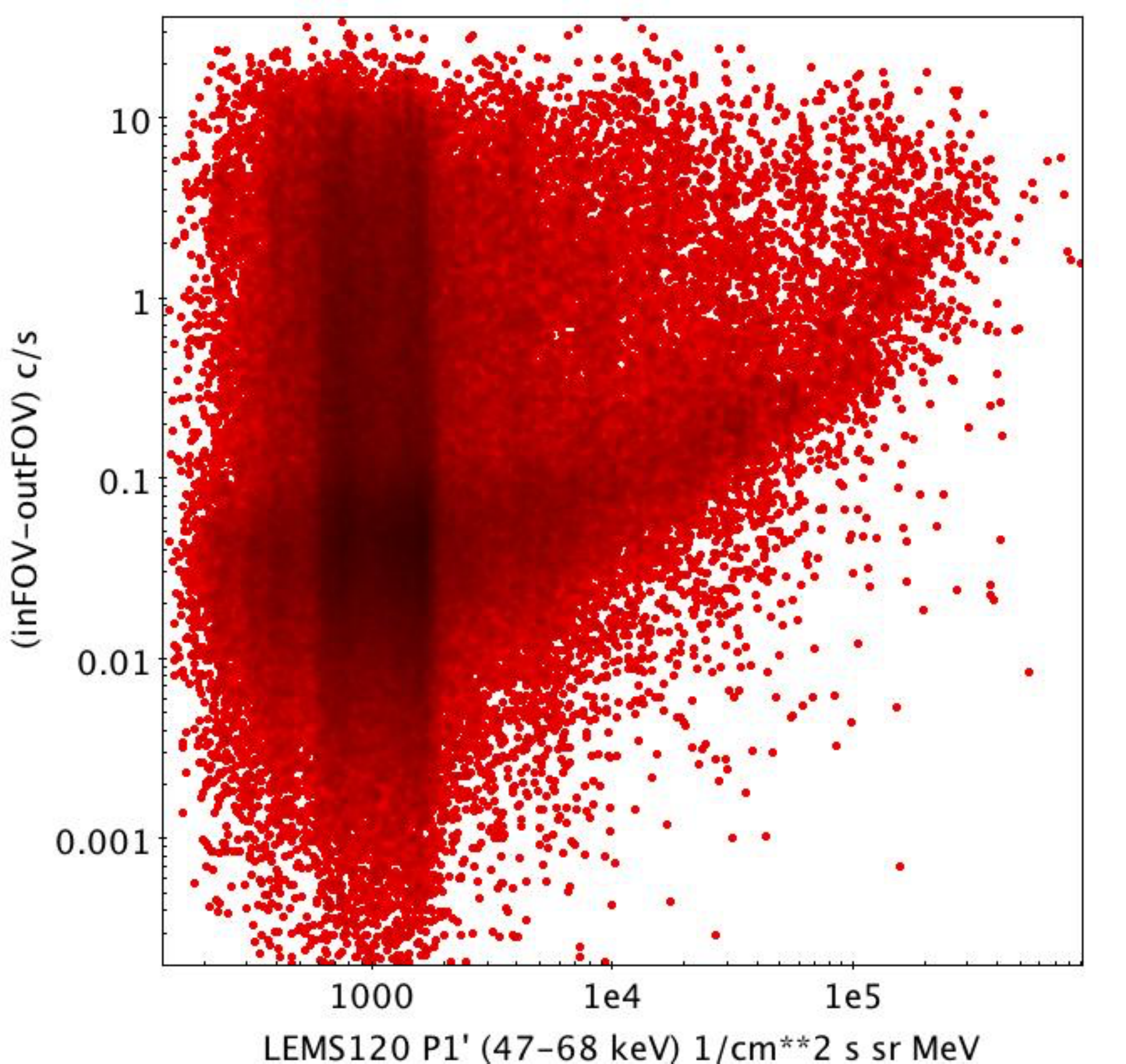}
  \includegraphics[width=0.5\textwidth]{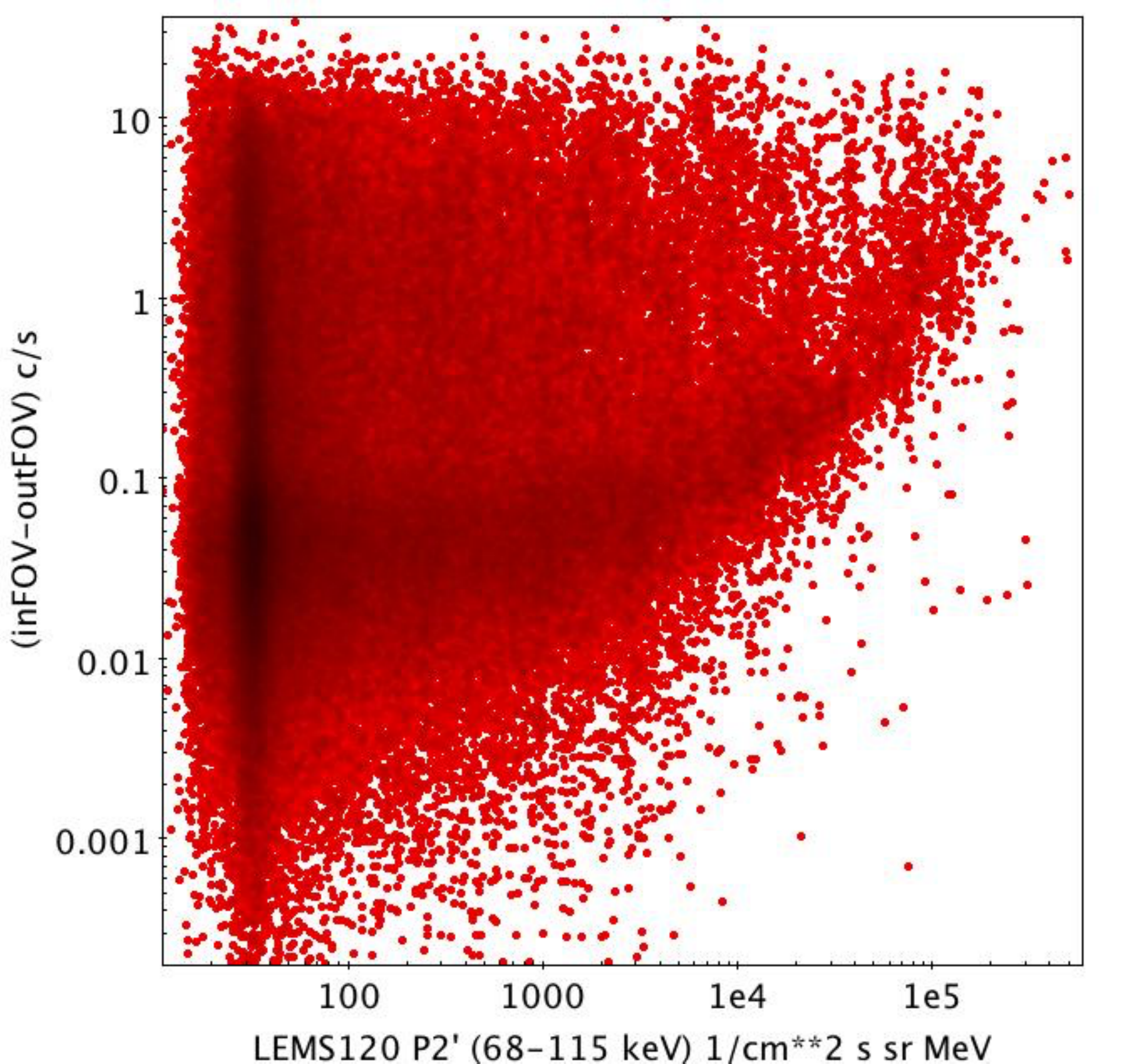}
\caption{Left Panel: Comparison of XMM-Newton \emph{inFOV}-\emph{outFOV} rates and ACE LEMS120 proton flux in the P1' channel (46-67 keV). Right Panel: Same as the left Panel but for the P2' channel (67-115 keV).}
\label{fig:9}       
\end{figure*}

We show the comparison of the EPIC MOS2 (\emph{inFOV}-\emph{outFOV}) rate and ACE LEMS120 proton flux
in the P1$^{\prime}$ and P2$^{\prime}$ channels in Figure \ref{fig:9}.
It is clear from the investigation of the plot that there is no striking correlation, besides a tendency
for a lower envelope, meaning that given a high flux of soft protons in L1 we can expect a corresponding
high level in EPIC. However at any given flux in L1 there is a wide range of intensities of soft protons
detected at the position of the XMM-Newton orbit, pointing to local (within the magnetosphere) acceleration sites 
for this particle component.
Much of the structure seen below 
$2\times10^{3}$/(cm$^{2}$ s sr MeV) in the P1$^{\prime}$ channel is due to 
background \cite{Haggerty.ea:06,Budjas.ea:17}.
The P2$^{\prime}$ channel is not affected by background problems and it provides the same basic picture.
We have not applied a delay time allowing for protons flight time from L1 to Earth, also because it is
not always clear the direction of travel (e.g. in the case of upstream events). We experimented applying 
delay times from 400s (the free streaming travel time from L1 to Earth for a 67 keV proton) up to 1hr
and the qualitative picture does not change.

\begin{figure*}
  \includegraphics[width=0.5\textwidth]{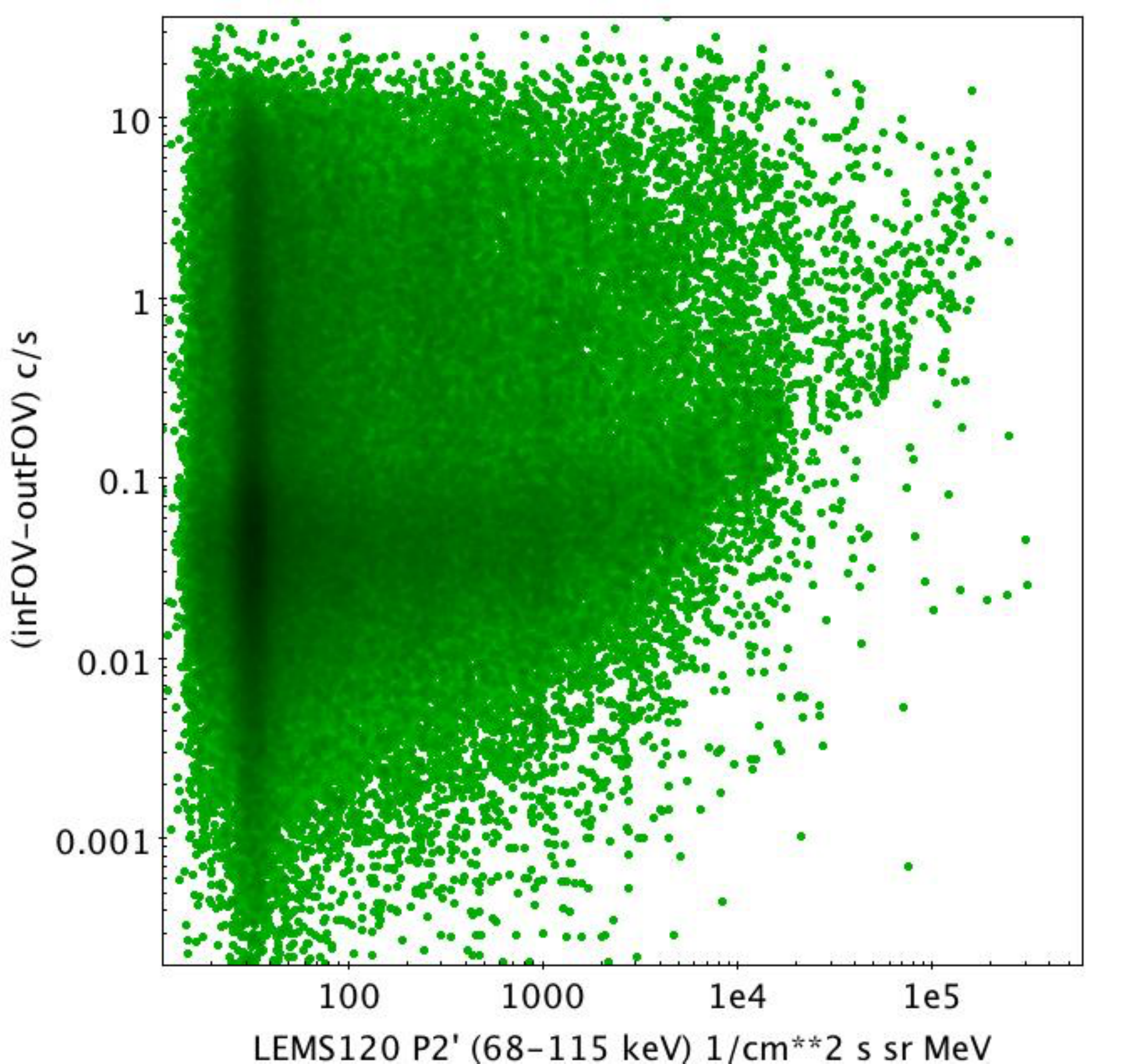}
  \includegraphics[width=0.5\textwidth]{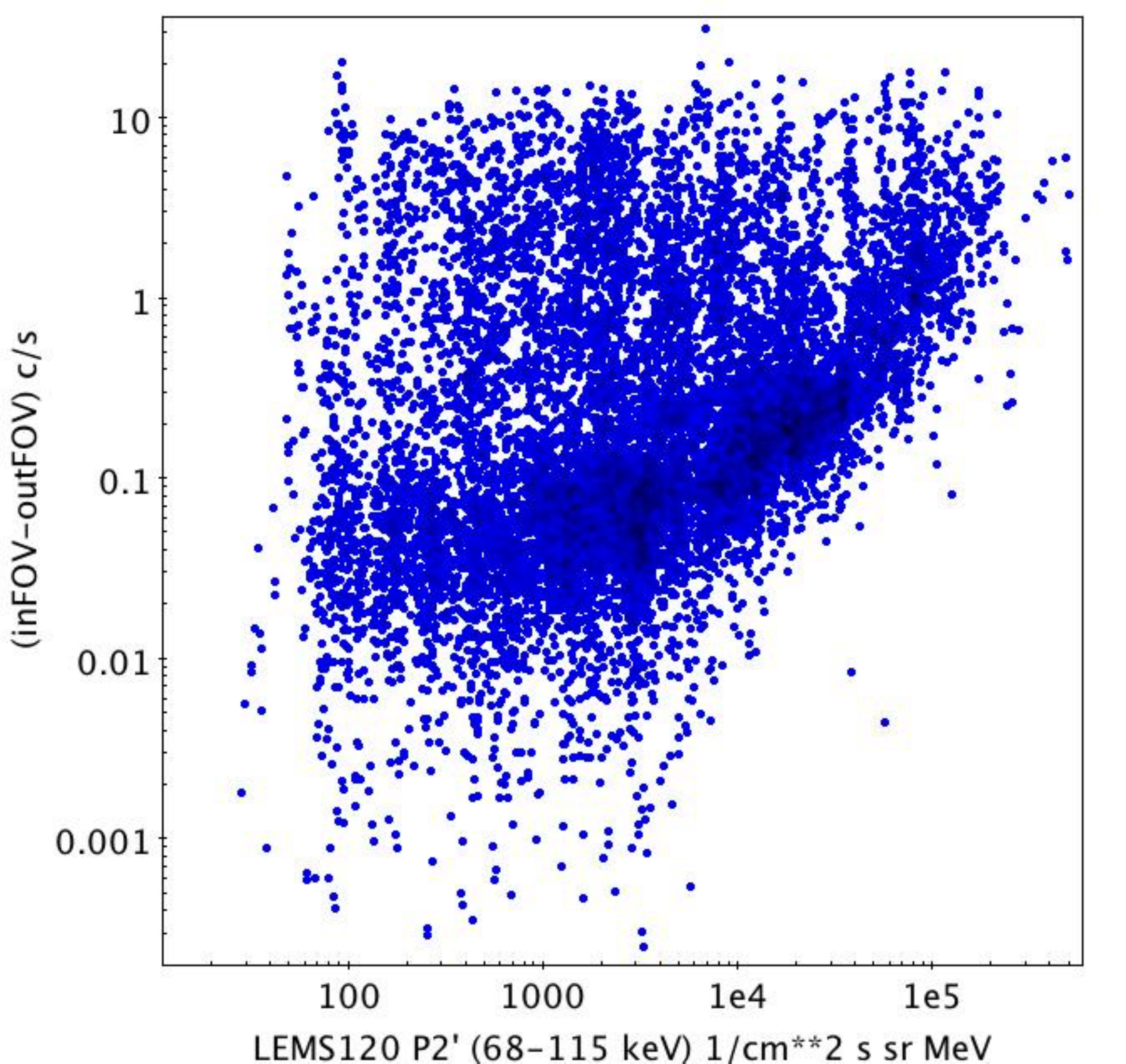}
\caption{Left Panel: Comparison of XMM-Newton \emph{inFOV}-\emph{outFOV} rates and ACE LEMS120 proton flux in the P2' channel (67-115 keV) during periods not affected by SEP events. Right Panel: Same as the left Panel but for periods during SEP events.}
\label{fig:10}       
\end{figure*}

If we divide our data when considering time intervals not affected by SEP events and time intervals
during SEP events (see Figure \ref{fig:10}) we can see that as expected the bulk of high proton fluxes
in L1 corresponds to SEP events, however this does not correspond to a better correlation in the EPIC
data. it is also to be noted that most of the time during SEP events EPIC is not observing to prevent
radiation damage.

\subsection{The \emph{inFOV}-\emph{outFOV} MOS2 and ACE EPAM LEMS data during SEPs}

Motivated by the non negligible amount of EPIC data obtained during SEP events, we investigated in detail 
the 92 SEP events occurred during the time span of our XMM-Newton data. We show in detail a SEP event 
during which the largest amount of data are available as an example of the general behavior.

\begin{figure}
  \includegraphics[width=1.0\textwidth]{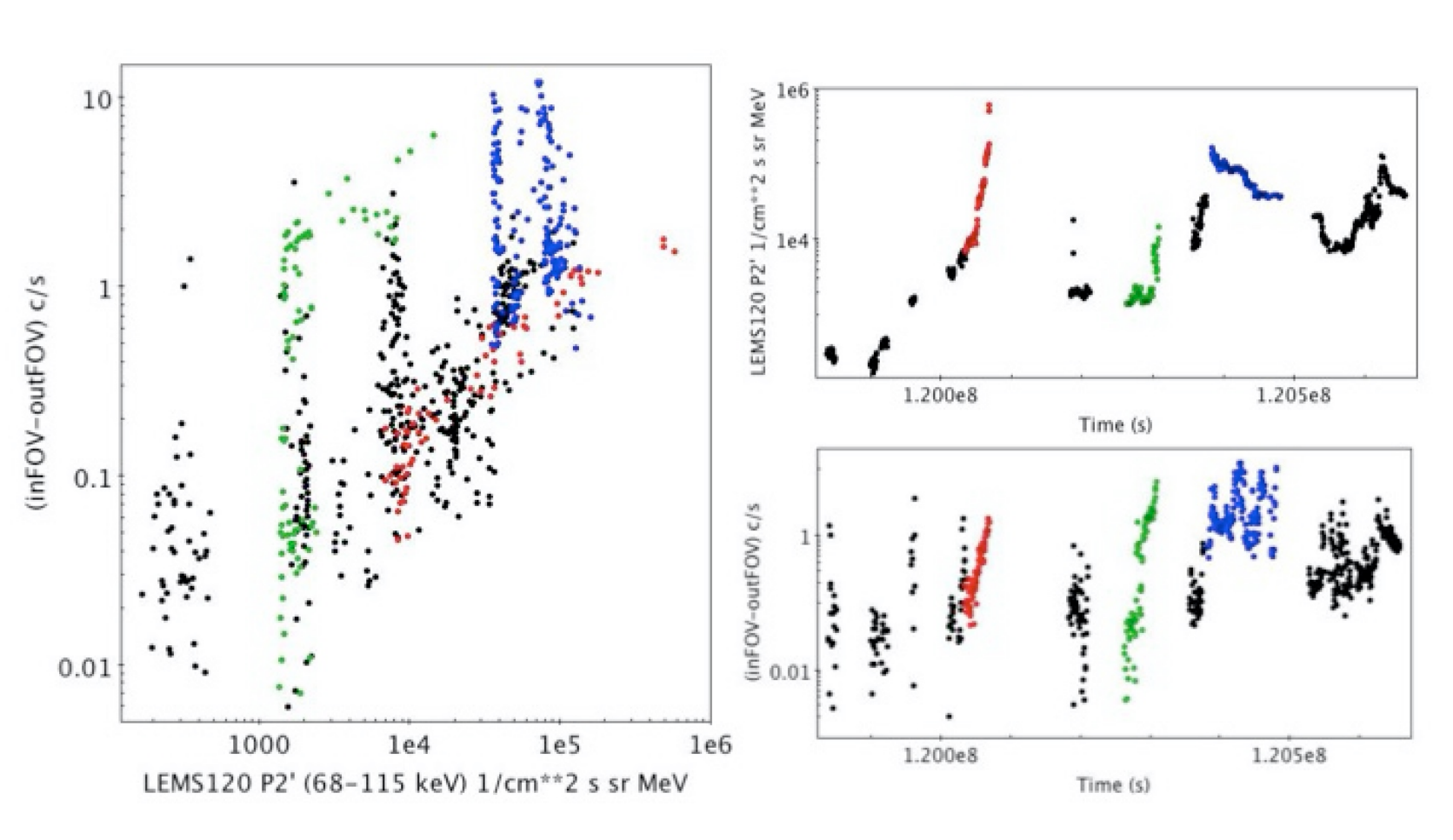}
\caption{EPIC MOS2 and ACE LEMS120 P2$^{\prime}$ data taken during the SEP event of 19-28 October 2001. Left Panel : Comparison of XMM (\emph{inFOV}-\emph{outFOV}) rates and ACE LEMS120 proton flux in the P2$^{\prime}$ channel (67-115 keV). Right upper panel: LEMS120 P2$^{\prime}$ light curve. Right bottom panel: EPIC MOS2 (\emph{inFOV}-\emph{outFOV}) light curve. Different parts of the light curves are depicted in different colors: red the portion showing a good correlation, green showing only a partial correlation, blue showing no correlation.}
\label{fig:11}       
\end{figure}

The case study shown in Figure \ref{fig:11} refers to the SEP event occurring in the time interval 19-28 
October 2001 where the amount of EPIC MOS2 data available are 387.5 ks. The plot of the comparison between 
EPIC MOS2 \emph{inFOV}-\emph{outFOV}  rate and ACE LEMS120 proton flux in the P2$^{\prime}$ channel shown in 
the left panel of Figure \ref{fig:11} shows the same qualitative trend of the one collecting all data during 
SEPs shown in the right 
panel of Figure \ref{fig:10}. Investigating in detail the light curves we highlighted different portions of 
them by 
different colors. If the part of the light curve painted in red shows a correlation, the one in green show a 
small correlation in the high MOS count rate part, whereas the one depicted in blue shows no correlation 
marking the "finger"-like structure well represented in the general plot of the right panel of Figure 
\ref{fig:10}.

\begin{figure}
  \includegraphics[width=1.0\textwidth]{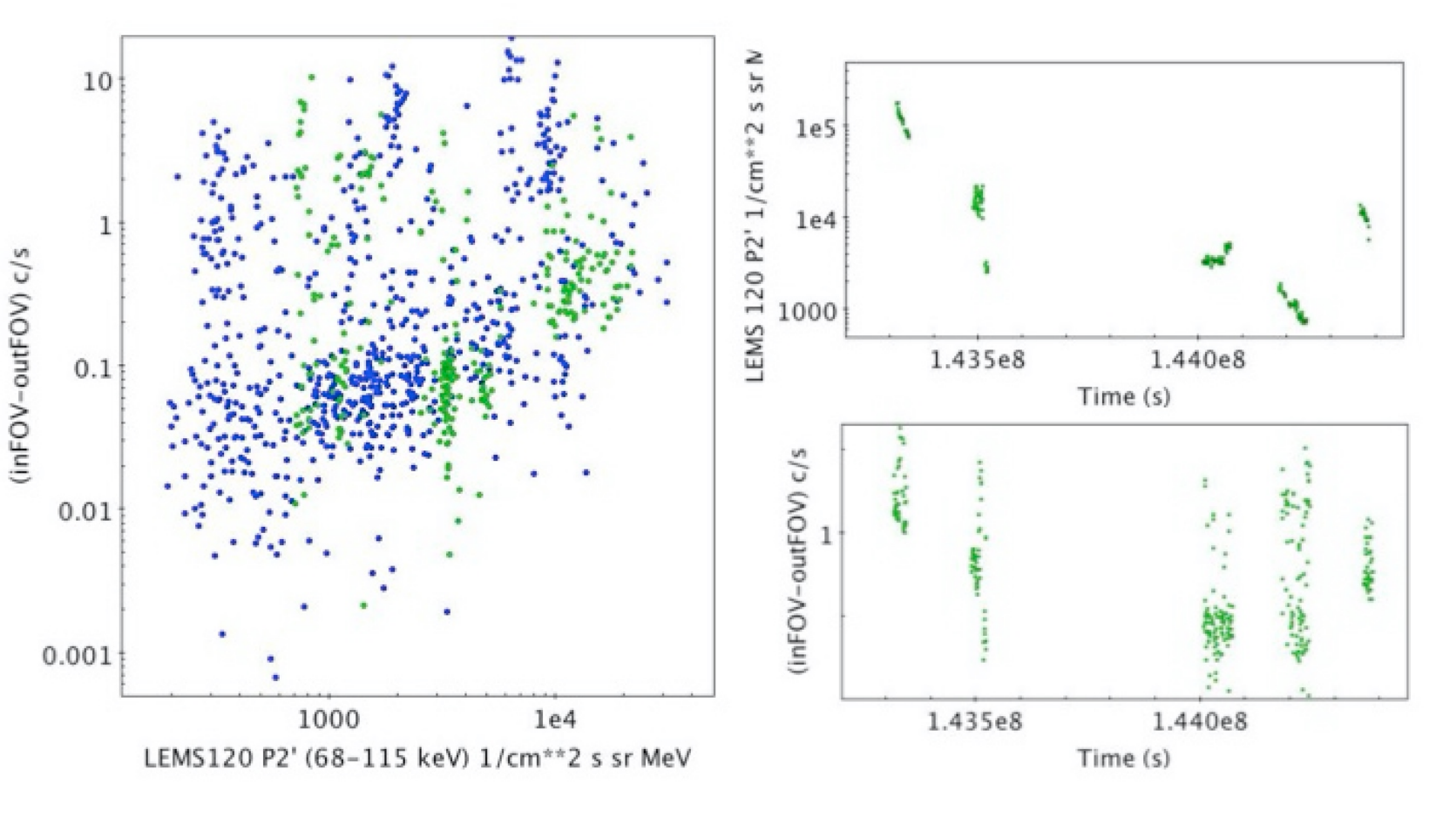}
\caption{Same as Figure \ref{fig:11} but for the EPIC MOS2 and ACE LEMS120 P2$^{\prime}$ data taken in time intervals affected by SEP events when XMM was outside of the bow shock. Highlighted in green and shown in the light curves are the data taken during the SEP period of 16-30 July 2002 for 165 ks.}
\label{fig:12}       
\end{figure}

In order to possibly disentangle the complication due to the propagation of protons in the magnetosphere we 
investigated the behavior of the two datasets when selecting time interval when a SEP event was ongoing and 
XMM was out of the bow shock. We found 534.5 ks of data satisfying the above conditions and spanning 13 
SEP events in the period from July 2000 to July 2005. The results are shown in Figure \ref{fig:12} with the 
same scheme as in the previous figure: despite the attempt of avoiding the complications due to the 
magnetosphere no clear trend emerged. This is an indication that the orientation of the satellite with 
respect to the local magnetic field plays possibly an important role.

\subsection{Comparison of  \emph{inFOV}-\emph{outFOV} MOS2 and ERM data}

\begin{figure}
  \includegraphics[width=0.65\textwidth]{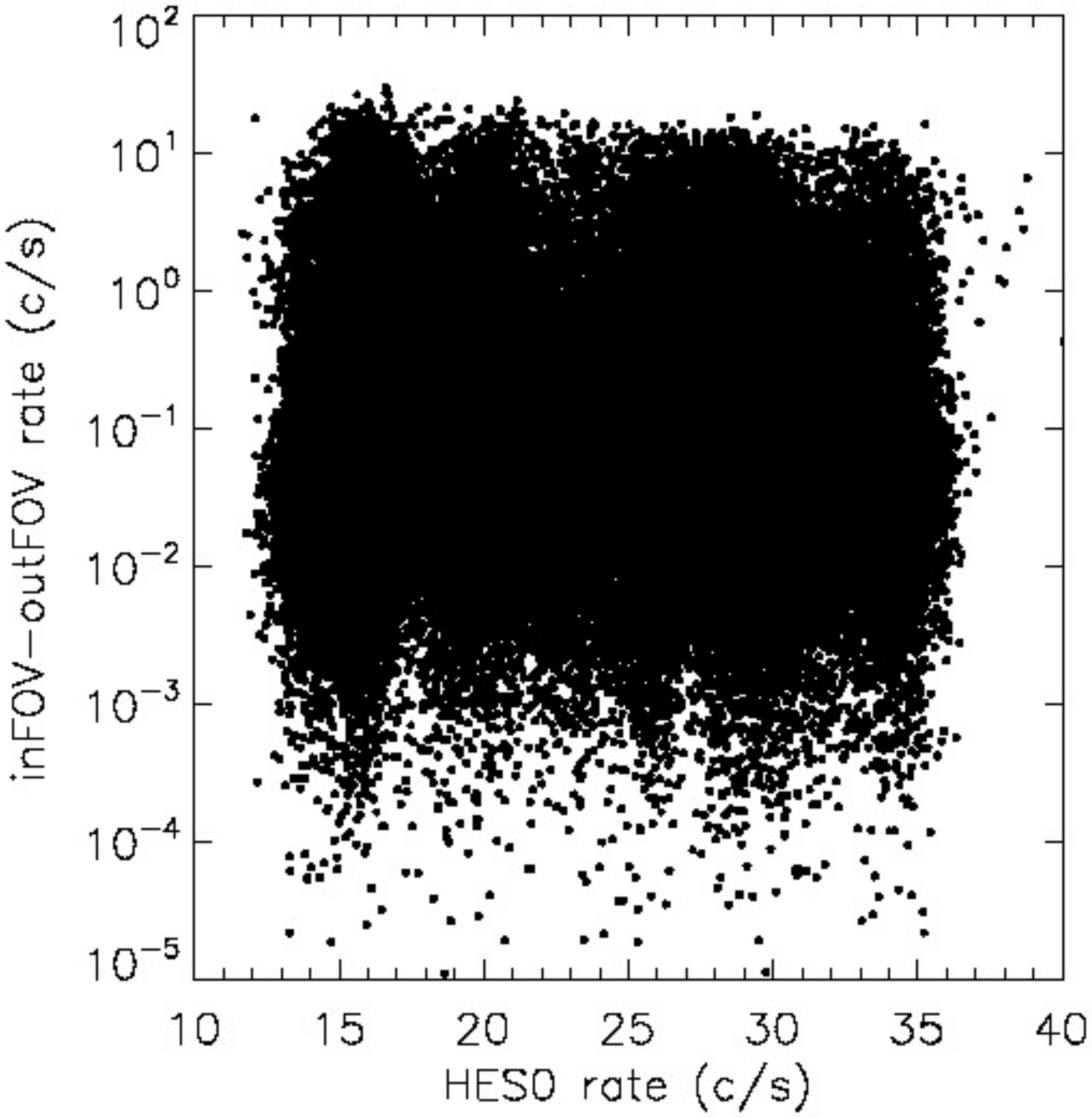}
\vskip -1.7truecm
\caption{Comparison between ERM HES0 count rates and the corresponding \emph{inFOV}-\emph{outFOV} count rate.}
\label{fig:13}       
\end{figure}

We also compared the ERM and \emph{inFOV}-\emph{outFOV} MOS2 data with the selection discussed
in Section \ref{sec:2.2}. The resulting plot (see Figure \ref{fig:13}) is strikingly different from the 
one presented in Section
\ref{sec:2.3} showing a clear lack of correlation, with a Spearman's $\rho$ of -0.07 and Kendall's $\tau$ 
of -0.048. This reinforces with the exquisite data statistics
of our project the fact that the focused soft proton component has energies below the one probed by the 
ERM as early recognized in the mission (e.g., \cite{Kendziorra.ea:00}).

\section{Summary and conclusions}

We have provided clear evidence that the XMM-Newton EPIC MOS2 instrumental background is clearly correlated 
with the flux of GCR, as modulated by the solar cycle. Correlation may not mean causation: 
relying on established understanding based on Geant 4 simulation the main element 
of the background are knock-on electrons ejected by the high energy GCR protons 
\cite{Hall.ea:10,Lotti.ea:14}. The minimization of this component for future detectors, in particular
the ones designed to fly on board Athena is actively pursued \cite{Lotti.ea:12,Lotti.ea:14}.

For what concern the focused particle background we are at an intermediate stage where strong conclusions
can not be reached yet, besides an indication of the large variety of acceleration sites for the soft 
protons. Clearly a measurement of the proton flux needs to be performed in a location as close as possible 
to the conditions experienced by XMM-Newton at that specific time.

%

\begin{acknowledgements}
The AHEAD project (grant agreement n. 654215) which is part of the EU-H2020 programm is acknowledged for 
partial support.
This work is part of the AREMBES WP1 activity funded by ESA through
contract No. 4000116655/16/NL/BW.
Results presented here are based, in part, upon work funded through the
European Union Seventh Framework Programme (FP7-SPACE-2013-1), under grant
agreement n. 607452, "Exploring the X-ray Transient and variable Sky -
EXTraS". We would like to thank Du{\v s}an Budj{\'a}{\v s} for a useful discussion about ACE data.
\end{acknowledgements}

\bibliographystyle{spphys}       
\bibliography{gasta.bib}   

%
%

\end{document}